\documentclass[24pt]{article}
\usepackage{amssymb}
\usepackage{amsfonts}
\usepackage{amssymb,amsmath}
\usepackage{mathrsfs}
\usepackage{latexsym}
\usepackage{amsmath}
\usepackage{latexsym}
\usepackage[cp1251]{inputenc}
\usepackage{graphicx}
\usepackage{color}

\title{Dilute Bose gas in classical environment at low temperatures}

\author{Volodymyr~Pastukhov\footnote{e-mail: volodyapastukhov@gmail.com}\\ {\small \textit{Ivan Franko
			National University of Lviv, Department for Theoretical Physics}}\\
	{\small \textit{12 Drahomanov St., Lviv, Ukraine }}}\date{}
\begin{document}
	\maketitle

	\begin{abstract}
		The properties of a dilute Bose gas with the non-Gaussian quenched disorder are analysed. Being more specific we have considered a system of bosons immersed in the classical bath consisting of the non-interacting particles with infinite mass. Making use of perturbation theory up to second order we have studied the impact of environment on the ground-state thermodynamic and superfluid characteristics of the Bose component.
	\end{abstract}

\section{Introduction}
The properties of Bose gas with quenched disorder was studied extensively during last two decades. This rise of an interest to such a system was stimulated by the possibility to observe Bose glass state transition \cite{Fisher} where superfluidity disappears \cite{Fisher_2} even at very low temperatures. The first attempts for microscopic description within the approximate second-quantization method adopted for Bose systems with disorder at low temperatures were undertaken in Ref.~\cite{Huang}. The further developments \cite{Muller_Gaul}, particularly the non-perturbative extensions on case of an arbitrary two-body coupling strength \cite{Yukalov} and strong external potential \cite{Navez,Yukalov_etal} generally confirm these findings. Diffusive Monte Carlo simulations \cite{Astrakharchik} also agree with the Bogoliubov-like result in the dilute limit, but the increase of disorder potential makes the differences more visible. The finite-temperature phase diagram of the system was clarified extensively in \cite{Kobayashi,Falco,Khellil}. A shift of the Bose-Einstein condensation transition temperature was determined in \cite{Zobay,Timmer} and the critical parameters of Bose gas in the disordered medium was calculated \cite{Pilati} using quantum Monte Carlo methods. No less interesting is the structure of quasiparticle spectrum and damping for a Bose system in the weak random external potential \cite{Giorgini,Gaul,Lugan,Lellouchl}. In particular, it was shown that the presence of disorder broadens the phonon peak of the dynamic structure factor which for a case of liquid $^4$He with the randomly distributed static impurities was computed \cite{Boninsegni} by means of Path Integral Monte Carlo method and experimentally measured in Ref.~\cite{Glyde}. Another consequence of disorder is the necessity to correct the exact universal identities which are characteristic for many-boson systems like the Hugenholtz-Pines theorem \cite{Lopatin} and Josephson's relation \cite{Muller}.

In recent experiments the disorder is usualy produced by the employment of an optical speckle potential, which characteristics are precisely controllable \cite{Semeghini}. The behavior of bosons in random potential created by laser speckles was also investigated theoretically \cite{Abdullaev,Astrakharchik_13,Fratini}. The simplest for understanding realization of disorder, however, can be achieved by the immersion of randomly distributed static impurities in the Bose condensate. When the concentration of impurity particles is small or the interaction with bosons is weak the distribution of the effective external potential appearing due to boson-impurity interaction can be modelled by the Gaussian. This is exactly the situation considered practically in all available theoretical studies concerning Bose gases with disorder. But for the diluted Bose systems the role of higher-order boson-impurity scattering processes increases that requires to go beyond the standard model of weak disorder and to include the distribution non-Gaussianity. The latter forms the main goal of present study.

\section{Formulation of problem}
We consider the system of $N$ interacting bosons immersed in the bath formed by $\mathcal{N}$ non-interacting classical (infinite mass) particles. This model is described by the following Euclidean action
\begin{eqnarray}\label{S}
S=\int dx \,\psi^*(x)\left\{\frac{\partial}{\partial\tau}+\frac{\hbar^2\nabla^2}{2m}+\mu
-\tilde{g}\rho({\bf r})\right\}\psi(x)-\frac{g}{2}\int dx|\psi(x)|^4,
\end{eqnarray}
where complex field $\psi(x)$ describes bosonic degrees of freedom, $\mu$ is the chemical potential that fixes the Bose gas density. The integration over $x=(\tau,{\bf r})$ is carried out in a $(3+1)$-domain of volume $\beta V$ ($\beta$ is the inverse temperature) with periodic boundary conditions. The quantity  $\rho({\bf r})=\sum_{1\ge j\le \mathcal{N}}\delta({\bf r}-{\bf r}_j)$ represents the density of homogeneously distributed classical particles. Both boson-boson and boson-impurity two-body interactions are assumed to be short-ranged that are characterised by the coupling constants $g$ and $\tilde{g}$, respectively. The latter should be related to the appropriate $s$-wave scattering lengths in the end of calculations
\begin{eqnarray}\label{g}
\frac{1}{g}=\frac{1}{t}-\frac{1}{V}\sum_{{\bf k}}\frac{1}{2\varepsilon_k}, \ \ \frac{1}{\tilde{g}}=\frac{1}{\tilde{t}}-\frac{1}{V}\sum_{{\bf k}}\frac{1}{\varepsilon_k},
\end{eqnarray}
where $t=4\pi\hbar^2a/m$, $\tilde{t}=2\pi\hbar^2\tilde{a}/m$ and $\varepsilon_k=\hbar^2k^2/2m$ is the free-particle dispersion. Introducing phase-density representation for bosonic fields $\psi^*(x)=\sqrt{n(x)}e^{-i\varphi(x)}$, $\psi(x)=\sqrt{n(x)}e^{i\varphi(x)}$ and making use of the Fourier transformation for $n(x)$ and $\varphi(x)$
\begin{eqnarray}\label{n_varphi}
n(x)=n+\frac{1}{\sqrt{\beta V}}\sum_{K}e^{iKx}n_{K}, \ \
\varphi(x)=\frac{1}{\sqrt{\beta V}}\sum_{K}e^{iKx}\varphi_{K},
\end{eqnarray}
where  $K=(\omega_k, {\bf k})$ stands for the bosonic Matsubara frequency $\omega_k$ and three-dimensional wave-vector ${\bf k}$, as well as for the density of classical component
\begin{eqnarray}
\rho({\bf r})=\rho+\frac{1}{\sqrt{V}}\sum_{{\bf k}\neq 0}e^{i{\bf kr}}\rho_{{\bf k}},
\end{eqnarray}
where $\rho=\mathcal{N}/V$ is average density of bath particles and $\rho_{\bf k}=\frac{1}{\sqrt{V}}\sum_{1\ge j\le \mathcal{N}}e^{i{\bf kr}_j}$, one rewrites action (\ref{S}) in the following way:
\begin{eqnarray}\label{SS}
S=S_B+S_d.
\end{eqnarray} 
The first term describes Bose gas itself \cite{Pastukhov_q2D}
\begin{eqnarray}\label{S_B}
S_B=\beta V\mu n-\frac{1}{2}\beta Vgn^2-\frac{1}{2}\sum_{K}
\left\{\vphantom{\left[\frac{\varepsilon_k}{2 n}
	+g\right]}\omega_k\varphi_K n_{-K}-\omega_k\varphi_{-K}n_{K}
\right.\nonumber\\
\left.+2n\varepsilon_k
\varphi_{K}\varphi_{-K}+\left[\frac{\varepsilon_k}{2 n}
+g\right]n_{K}n_{-K}\right\}\nonumber\\
+\frac{1}{3!\sqrt{\beta
		 V}}\sum_{K+Q+P=0}\frac{1}{4n^2}(\varepsilon_k+\varepsilon_q+\varepsilon_p)n_{K}n_{Q}n_{P}\nonumber\\
+\frac{1}{2\sqrt{\beta V}}\sum_{K,
	Q}\frac{\hbar^2}{m}{\bf
	kq}\varphi_{K}\varphi_{Q}n_{-K-Q} + \ldots,
\end{eqnarray}
while $S_d$ takes into account the presence of environment
\begin{eqnarray}\label{S_d}
S_d=-\beta V\tilde{g}n\rho-\sqrt{\beta}\tilde{g}\sum_{K}\delta_{\omega_k,0}\rho_{-{\bf k}}
n_{K}.
\end{eqnarray}
The thermodynamic relation $-\partial \Omega/\partial\mu=N$ for the grand potential together with explicit form of Eq.~(\ref{S_B}) fix $n=N/V$ \cite{Pastukhov_15,Pastukhov_InfraredStr} to be the density of the Bose system. This observation allows to proceed in the canonical ensemble and in order to obtain physically meaningful result the averaging over the positions of classical particles should be performed for the free energy of our system. It particularly means that first we have to calculate the free energy $F$ of Bose gas in the presence of local external potential $\tilde{g}\rho({\bf r})$ and then 
identify the Helmholtz potential of the system ``Bose gas + classical bath'' with 
\begin{eqnarray}\label{bar_F}
\bar{F}=\frac{1}{V^{\mathcal{N}}}\int_V d{\bf r}_1\ldots\int_V d{\bf r}_{\mathcal{N}}F.
\end{eqnarray}

In the following we will assume that bosons are weakly-coupled to each other and study the impact of the classical component on the properties of Bose gas in the zero-temperature limit. It is well-known that presence of disorder or interaction with other quantum systems depletes the superfluid density of the Bose gas even at absolute zero \cite{Andreev,Viverit,Pastukhov_twocomp,Utesov}. Of course, this phenomena is observed in our case too, and in order to calculate the normal density of superfluid we have to assume that the Bose subsystem moves as a whole with velocity ${\bf v}$. Simple analysis \cite{Konietin} shows that account of this motion only leads to the shift $\omega_k\to\omega_k-i\hbar{\bf vk}$ of Matsubara frequency in action $S_B$. Then the general consideration in the spirit of perturbation theory in terms of $\tilde{g}$ leads to
\begin{eqnarray}\label{E}
\bar{E}_{{\bf v}}=E_B+N\frac{m{\bf v}^2}{2}+V\tilde{g}n\rho-\frac{1}{2}\sum_{{\bf k}}\tilde{g}^2
 \overline{\rho_{{\bf k}}\rho_{-{\bf k}}}\langle n_{K}n_{-K}\rangle|_{\omega_k=0}
\nonumber\\
+\frac{1}{3!}\sum_{{\bf k}+{\bf q}+{\bf p}=0}\tilde{g}^3\overline{\rho_{{\bf k}}\rho_{{\bf q}}\rho_{{\bf p}}}\sqrt{\beta}\langle n_{K}n_{Q}n_{P}\rangle|_{\omega_k=\omega_q=\omega_p=0} + \ldots,
\end{eqnarray}
for the ground-state energy of system, where $E_B$ is the contribution of the Bose gas in rest alone and $ \langle n_{K}n_{-K}\rangle$, $\langle n_{K}n_{Q}n_{P}\rangle,\ldots$ denote the irreducible density correlation functions of pure bosons moving with velocity ${\bf v}$. The structure factors of classical non-interacting particles are fully determined by their density and can be easily evaluated by using the procedure described above: $\overline{\rho_{{\bf k}}\rho_{-{\bf k}}}=\rho$, $\overline{\rho_{{\bf k}}\rho_{{\bf q}}\rho_{{\bf p}}}=\rho\delta_{{\bf k}+{\bf q}+{\bf p},0}/\sqrt{V}$, etc. Expanding r.h.s. of Eq.~(\ref{E}) in powers of velocity
\begin{eqnarray}
\bar{E}_{{\bf v}}=\bar{E}_{{\bf v}=0}+V\frac{m{\bf v}^2}{2}n_s+o({\bf v}^2),
\end{eqnarray}
(actually in powers of dimensionless parameter $v/c$, where $c$ is the sound velocity) to quadratic order one obtains the density $n_s$ of superfluid component.

In addition to the depletion of superfluid density the interaction with bath also decreases the number of Bose particles with zero momentum. To calculate the condensate density of Bose gas we use the following prescription: first within variational differentiation
\begin{eqnarray}\label{N_k}
N_k=\left(\frac{\delta \bar{E}}{\delta \varepsilon_k}\right)_{n},
\end{eqnarray}
we determine the distribution function of particles with non-zero momentum and then obtain the condensation fraction 
\begin{eqnarray}\label{N_k}
\frac{n_0}{n}=1-\frac{1}{N}\sum_{{\bf k}\neq 0}N_k.
\end{eqnarray}
To this stage our consideration is formally exact and the problem is actually reduced to the calculation of irreducible density correlators of a pure Bose gas. But even in the Bogoliubov approximation these calculations are very cumbersome, therefore below we restrict ourselves to the case of weak interspecies interaction, where the perturbation theory in terms of $\tilde{g}$ can be used.

\section{Perturbation theory}
By treating the Bose subsystem on the basis of Bogoliubov's theory we greatly simplify the analysis below but at that time restrict it to the consideration of dilute gases. This is exactly the situation realized in experiments with cold alkali atoms. From the point of view of further calculations in the dilute limit we are free to drop the beyond Bogoliubov corrections to the various density correlation functions. Additionally, working with the same accuracy one should treat the ground-state energy of pure bosons $E_B=E_{LHY}$ on level of the Lee-Huang-Yang \cite{Lee_Huang_Yang} formula.

Following the above-mentioned approximation scheme in the first order of perturbation theory we have to neglect the last term in Eq.~(\ref{E}) and substitute the pair density correlation function 
\begin{eqnarray}
\langle n_{K}n_{-K}\rangle=\frac{2n\varepsilon_k}{E^2_k+(\omega_k-i\hbar{\bf k}{\bf v})^2},
\end{eqnarray}
with $E_k=\sqrt{\varepsilon^2_k+2ng\varepsilon_k}$ being Bogoliubov's spectrum. The resulting formula has to be used for obtaining the particle distribution $N_k$ and condensate density
\begin{eqnarray}
\frac{n_0}{n}=\frac{n^{B}_0}{n}-\frac{1}{V}\sum_{{\bf k}\neq 0}\rho{\tilde{t}}^2\frac{\varepsilon^2_k}{E^4_k},
\end{eqnarray}
($n^{B}_0/n
$ is the Bogoliubov result for pure bosons) in the adopted approximation, and after renormalization of the coupling constant (\ref{g}) for the explicit evaluation of the energy correction 
\begin{eqnarray}
	\frac{\bar{E}^{(1)}_{{\bf v}=0}}{N}=\rho\tilde{t}-\frac{1}{V}\sum_{{\bf k}\neq 0}\rho{\tilde{t}}^2\left[\frac{\varepsilon_k}{E^2_k}-\frac{1}{\varepsilon_k}\right],
\end{eqnarray}
and depletion of the superfluid component
\begin{eqnarray}
\frac{n_s}{n}=1-\frac{4}{3V}\sum_{{\bf k}\neq 0}\rho{\tilde{t}}^2\frac{\varepsilon^2_k}{E^4_k}.
\end{eqnarray}
From general principles as well as from the above formulas it is clear that in the first-order approximation the results derived for our system are identical to those obtained for a dilute Bose gas with weak disorder \cite{Astrakharchik}. The differences appear in the next orders of the formulated perturbation theory.

The second order calculations require the account of the last term in Eq.~(\ref{E}). The appropriate three-point density correlator reads
\begin{eqnarray}
\langle n_{K}n_{Q}n_{P}\rangle=\frac{\delta_{K+Q+P,0}}{\sqrt{\beta V}}\left[\frac{\hbar^2}{m}{\bf k}{\bf q}\langle n_{K}\varphi_{-K}\rangle\langle n_{Q}\varphi_{-Q}\rangle\langle n_{P}n_{-P}\rangle\right.\nonumber\\
\left.+\frac{\varepsilon_k}{4n^2}\langle n_{K}n_{-K}\rangle\langle n_{Q}n_{-Q}\rangle\langle n_{P}n_{-P}\rangle+\textrm{perm.}\right],
\end{eqnarray}
in the dilute limit, where we have used shorthand notation for the phase-density correlator
\begin{eqnarray}
\langle \varphi_{K}n_{-K}\rangle=\frac{\omega_k-i\hbar{\bf k}{\bf v}}{E^2_k+(\omega_k-i\hbar{\bf k}{\bf v})^2}.
\end{eqnarray}
The further strategy is the same as previously used. By calculating the next correction to the particle distribution $N_k$ we are in position to obtain the condensate density up to the second-order of perturbation theory. Then the application of the coupling constant renormalization procedure (\ref{g}) yields
\begin{eqnarray}
\frac{6}{V^2}\sum_{{\bf k},{\bf q}\neq 0}\rho\tilde{t}^3nt\frac{\varepsilon_k}{E^2_k}\frac{\varepsilon_q}{E^2_q}\frac{\varepsilon^2_{|{\bf k}+{\bf q}|}}{E^4_{|{\bf k}+{\bf q}|}}-\frac{2}{V}\sum_{{\bf k}\neq 0}\rho\tilde{t}^2\frac{\varepsilon^2_k}{E^4_k}\frac{1}{V}\sum_{{\bf q}\neq 0}\tilde{t}\left[\frac{\varepsilon_q}{E^2_q}-\frac{1}{\varepsilon_q}\right],
\end{eqnarray}
for the fraction of non-condensed particles. In the same fashion we obtain the energy correction
\begin{eqnarray}
	\frac{\bar{E}^{(2)}_{{\bf v}=0}}{N}=\frac{1}{V^2}\sum_{{\bf k},{\bf q}}\rho\tilde{t}^3\left\{\frac{\varepsilon_k}{E^2_k}\frac{\varepsilon_q}{E^2_q}
	\left[\frac{\varepsilon^2_{|{\bf k}+{\bf q}|}}{E^2_{|{\bf k}+{\bf q}|}}-1\right]+\left[\frac{\varepsilon_k}{E^2_k}-\frac{1}{\varepsilon_k}\right]\left[\frac{\varepsilon_q}{E^2_q}-\frac{1}{\varepsilon_q}\right]  \right\},
\end{eqnarray}
and the second-order normal density fraction
\begin{eqnarray}
\frac{2}{3V^2}\sum_{{\bf k},{\bf q}}\rho\tilde{t}^3\left(\frac{\hbar^2 {\bf k}{\bf q}}{m}\right)^2\frac{\varepsilon_{|{\bf k}+{\bf q}|}}{E^2_kE^2_qE^2_{|{\bf k}+{\bf q}|}}
-\frac{4}{3V^2}\sum_{{\bf k},{\bf q}}\rho\tilde{t}^3
\frac{\varepsilon^3_{|{\bf k}+{\bf q}|}}{E^4_{|{\bf k}+{\bf q}|}}\frac{\varepsilon_q}{E^2_q}\frac{\varepsilon_k}{E^2_k}\nonumber\\
-\frac{8}{3V^2}\sum_{{\bf k},{\bf q}}\rho\tilde{t}^3\left\{\frac{\varepsilon^2_k}{E^4_k}
\left[\frac{\varepsilon^2_q}{E^2_q}-1\right]\frac{\varepsilon_{|{\bf k}+{\bf q}|}}{E^2_{|{\bf k}+{\bf q}|}}+
\frac{\varepsilon^2_k}{E^4_k}\left[\frac{\varepsilon_q}{E^2_q}-\frac{1}{\varepsilon_q}\right]\right\},
\end{eqnarray}
associated with the presence of disorder.

\section{Results}
Interesting feature of the model with a delta-like potential is that all integrals appearing during the calculation of energy corrections, condensate and superfluid densities depletion can be evaluated analytically to the very end. Particularly for the fraction of condensed particles in the presence of classical bath we obtained the following expansion:
\begin{eqnarray}
\frac{n_0}{n}=\frac{n^{B}_0}{n}-\frac{\sqrt{\pi}}{2}\frac{\rho\tilde{a}^2}{\sqrt{na}}
-6\pi\rho\tilde{a}^3.
\end{eqnarray}
After tedious integration an expression for a superfluid density with a similar structure was also derived
\begin{eqnarray}
\frac{n_s}{n}=1-\frac{2\sqrt{\pi}}{3}\frac{\rho\tilde{a}^2}{\sqrt{na}}
-4\pi\left\{3-\frac{2}{3}\ln\frac{16}{3}\right\}\rho\tilde{a}^3.
\end{eqnarray}
The situation with energy is more complicated. There is no problem in calculation of the first-order correction, but the second one is logarithmically divergent. Let us remind that the same type of problem originally occurs \cite{Hugenholtz} during the computation of the beyond Lee-Huang-Yang ground-state energy of pure Bose gas and is totally connected with the point-like approximation of the two-body potential. Applying similar regularization procedure, i.e., cutting off the upper integration limit on a scale of order $1/\tilde{a}$ we finally have (with logarithmic accuracy)
\begin{eqnarray}
\frac{\bar{E}_{{\bf v}=0}}{N}=\frac{E_{LHY}}{N}+\rho\tilde{t}
\left\{1+4\sqrt{\pi}\sqrt{na\tilde{a}^2}-8\pi na\tilde{a}^2\ln\frac{1}{na\tilde{a}^2} \right\}.
\end{eqnarray}
Of course, this result can be justified within more sophisticated consideration which particularly intends the explicit momentum dependence of the $t$-matrix (\ref{g}). Indeed, in the limit of $k\to \infty$ the leading asymptote is the following $\tilde{t}\sim 1/(k\tilde{a})^2$ {in the limit of vanishing effective interaction range}  that provides the convergence of integrals and correctness of the above cut-off procedure.

\section{Conclusions}
In summary, by means of hydrodynamic approach we have studied properties of a dilute Bose gas with the non-Gaussian quenched disorder. The realization of external random potential is performed by inserting into the system a macroscopic number of non-interacting classical particles with infinite mass. Assuming that the two-body potential describing interaction between bath particles and bosons is short-ranged and weak we have perturbatively analyzed the thermodynamic and superfluid characteristics of the system. Particularly  we obtained, in addition to the well-known Bogoliubov-like result, the second-order beyond-mean-field corrections to the energy, condensate fraction and superfluid density of Bose gas. It is instructive to note that the presence of environment generally depletes the superfluid and condensate densities, furthermore, the second-order terms of these observables do not depend on number of bosons and are totally determined by the interaction with impurities. 

The possible experimental visualization of calculated the next to beyond-mean-field effects can be realised not only on the system of ``dirty'' bosons. Very promising in this context is a two-component mixture of Fermi particles \cite{Navon}, where the strength of the tunnable interaction can be tuned in wide range to observe both the weakly non-ideal Fermi gas and the dilute Bose condensate of dimers.

\section*{Acknowledgements}
We thank Prof.~A.~Rovenchak for invaluable comments. This work was partly supported by Project FF-30F (No.~0116U001539) from the Ministry of Education and Science of Ukraine.

\newpage

\end{document}